\documentclass[proof]{WileyASNA-v1_MC}

\articletype{ORIGINAL ARTICLE}

\received{\today}
\revised{}
\accepted{}

\usepackage[rightcaption]{sidecap}
\hypersetup{
    final=true,
    pageanchor=true,
    colorlinks=true,
    breaklinks=true,
    linkcolor=blue,
    citecolor=blue,
    urlcolor=blue,
    pdfpagemode=UseNone,
    pdftitle={},
    pdfauthor={},
    pdfsubject={},
    pdfkeywords={}}

\usepackage{mathptmx}

\usepackage[modulo, switch]{lineno}
\newcommand\arcsec{\mbox{$^{\prime\prime}$}}

\newcommand{\ns}{\hspace*{-5pt}}

\begin{document}


\title{Cloud model inversions of strong chromospheric absorption lines using
    principal component analysis}

\author[1,2]{E.\ Dineva*}
\author[1]{M.\ Verma}
\author[3]{S.J.\ Gonz\'alez Manrique}
\author[3]{P.\ Schwartz}
\author[1]{C.\ Denker}

\authormark{E.\ Dineva \textsc{et al.}}

\address[1]{\orgname{Leibniz-Institut f\"{u}r Astrophysik Potsdam (AIP)}, 
    \orgaddress{\city{Potsdam}, 
    \country{Germany}}}

\address[2]{\orgname{Universit{\"a}t Potsdam},
    \orgdiv{Institut f{\"u}r Physik und Astronomie}, 
    \orgaddress{\city{Potsdam}, 
    \country{Germany}}}

\address[3]{\orgname{Slovak Academy of Sciences}, 
    \orgdiv{Astronomical Institute}, 
    \orgaddress{\city{Tatransk\'a, Lomnica}, 
    \country{Slovak Republic}}}

\corres{*\email{edineva@aip.de}}

\presentaddress{Ekaterina Dineva, 
    Leibniz-Institut f\"{u}r Astrophysik Potsdam (AIP),
    An der Sternwarte~16,
    14482 Potsdam,
    Germany}

\jnlcitation{\cname{%
\author{E.\ Dineva}, 
\author{M.\ Verma}, 
\author{S.J.\ Gonz\'alez Manrique},
\author{P.\ Schwartz} and 
\author{C.\ Denker}} (\cyear{2019}), 
\ctitle{Cloud model inversions of strong chromospheric absorption lines using
    principal component analysis}, 
\cjournal{ASNA}, \cvol{}.}


\abstract[Abstract]{High-resolution spectroscopy of strong chromospheric 
absorption lines delivers nowadays several millions of spectra per observing 
day, when using fast scanning devices to cover large regions on the solar 
surface. Therefore, fast and robust inversion schemes are needed to explore the 
large data volume. Cloud Model (CM) inversions of the chromospheric H$\alpha$ 
line are commonly employed to investigate various solar features including 
filaments, prominences, surges, jets, mottles, and (macro-)spicules. The choice 
of the CM was governed by its intuitive description of complex chromospheric 
structures as clouds suspended above the solar surface by magnetic fields. 
This study is based on observations of active region NOAA~11126 in H$\alpha$, 
which were obtained 2010 November 18\,--\,23 with the echelle spectrograph of 
the Vacuum Tower Telescope (VTT) at the Observatorio del Teide, Spain. Principal 
Component Analysis (PCA) reduces the dimensionality of spectra and conditions 
noise-stripped spectra for CM inversions. Modeled H$\alpha$ intensity and contrast 
profiles as well as CM parameters are collected in a database, which facilitates 
efficient processing of the observed spectra. Physical maps are computed 
representing the line-core and continuum intensity, absolute contrast, equivalent 
width, and Doppler velocities, among others. Noise-free spectra expedite the 
analysis of bisectors. The data processing is evaluated in the context of 
``big  data'', in particular with respect to automatic classification of 
spectra.}


\maketitle


\section{Introduction}\label{SEC1}

Principal Component Analysis \citep[PCA, as an introductory manual to PCA see,  
e.g., ][]{Jolliffe2002} is a statistical technique to identify patterns in 
large multidimensional datasets. The rapid progress in technology provoked the 
need to cope with large data volumes driven by the fast acquisition of 
high-resolution data \citep{Zhang2015, Denker2018}. In turn, this demands fast 
analytic tools, which produce sensible results. In this context, PCA is a 
convenient tool for dimensionality reduction, noise-stripping, and highlighting 
data patterns, thus providing insight revealing spectral signatures and 
information such as line-wing and line-core emissions, line asymmetries, 
and multi-lobed spectra.

In astrophysics, PCA was applied to calibration, analysis, and classification of 
stellar, galactic, and interstellar spectra \citep{Brunt2013}, where it is often 
prerequisite for other machine learning techniques \citep{Kuntzer2016}. This 
approach uses the fact that certain physical properties of a source result in 
absorption spectra of similar shape. Thus, the classification problem simplifies 
to recognizing patterns in the spectral profile's shape.

\citet{Rees2000} presented PCA as a faster solution for spectral inversions as 
compared to the most commonly used trial-and-error methods based on non-linear 
least-squares fitting. Instead of computationally intense matching of every 
model profile to the observed one, they take advantage of PCA's pattern 
recognition potential. PCA is favored as a method for dimensionality reduction 
and an alternative to, for example, Fourier decomposition. The authors evaluate the 
compatibility of PCA with respect to various complex models of photospheric and 
chromospheric line formation.

In particular, \citet{Rees2000} discussed applications of PCA to the strong chromospheric 
absorption line H$\alpha$ $\lambda$6562.8~\AA\ and to spectropolarimetric 
observations of the photospheric Fe\,\textsc{i} $\lambda$5250.3~\AA\ line. This 
work with data of the Advanced Stokes Polarimeter \citep[ASP,][]{Elmore1992} at 
the Dunn Solar Telescope (DST), Sunspot, New Mexico was expanded by 
\citet{Socas_Navarro2001} who observed the pair of commonly used Fe\,\textsc{i} 
lines at $\lambda$6302~\AA. The question is raised whether and how it is 
possible to construct an adequate synthetic database \citep{LopezAriste2001}, 
which will not lose its accurate physical foundation by using PCA compression. 
The Fast Analysis Technique for the Inversion of Magnetic Atmospheres 
\citep[FATIMA,][]{Socas_Navarro2001} is an answer to this challenge and
implements a fast yet reliable inversion algorithm.

The benefits of PCA for high-resolution spectroscopy, and specifically for noise 
reduction, are discussed in \citet{Chae2013}. New instrumentation specifications 
and data reduction methods implementing PCA compression are put forward for the 
Fast Imaging Solar Spectrograph (FISS) installed at the 1.6-meter Goode Solar 
Telescope \citep[GST,][]{Goode2003a} at Big Bear Solar Observatory (BBSO).

Noise-free spectra potentially enhance the robustness and precision
of spectral inversion techniques.
In its classical form, the Cloud Model (CM) of \citet{Beckers1964},
is an attempt to distinguish and evaluate physical properties of dynamic 
chromospheric fine-structure in comparison with the undisturbed quiet-Sun 
background. The initial plasma parameters are assumed to be constant throughout 
the observed absorption feature, even though refined CMs drop this limitation. 
\citet{Tziotziou2007} provided a comprehensive summary, including examples, of 
the CM and its modifications, and discussed their implications for inversion 
techniques.

Observations in the Balmer H$\alpha$ line reveal an abundance of information 
about the fine structure of the solar chromosphere. The diagnostic power of 
H$\alpha$ spectroscopy results from the large span of the core-to-wing formation 
height. The wings are formed with photons arising from the photosphere, while 
the line-core originates in the chromosphere. This provides a tomographic view
across several scale heights in the solar atmosphere. As an example, 
\citet{Kuckein2016} studied the 
dynamical evolution and physical properties of a giant filament and compared 
them to smaller analogues. They used high-resolution echelle spectra in two 
different wavelength ranges (H$\alpha$ and Na\,\textsc{i}\,D$_2$ 
$\lambda$5890~\AA). Counter-streaming flows were detected in the filament 
structure based on velocities derived with CM inversions and Doppler shifts from 
least-squares parabola fitting of the line cores.

CM inversions of H$\alpha$ contrast profiles were presented by 
\citet{GonzalezManrique2017} who used least-squares minimization 
\citep{Markwardt2009} to infer the CM parameters. Their study focused on a 
small-scale arch filament system (AFS) in an emerging flux region (EFR). The 
observed region on the Sun contained two micro-pores with sizes slightly larger 
than one second of arc. In summary, CM inversion is a versatile tool 
investigating chromospheric activity from the arcsecond-scale flux emergence to 
large-scale filaments encircling the Sun.

This work is motivated by past and recent observations with the echelle 
spectrograph of the 0.7-meter Vacuum Tower Telescope 
\citep[VTT,][]{vonderLuehe1998} at the Observatorio del Teide, Tenerife, Spain. 
Many observing campaigns over the past decade produced spatio-spectral data cubes 
containing strong chromospheric absorption lines and a variety of solar 
features. In the era of ``big data'', our goal is to uniformly process these 
data and use them for database research \citep{Denker2019}, extracting 
information using machine learning for feature identification and spectral 
classification.

The various steps of data processing and analysis provide the structure for 
organizing our investigation. In Sect.~\ref{SEC2}, we discuss the observational 
settings and the standard data reduction procedures, i.e., (pre)processing of the 
spatio-spectral data cube, determining the quiet-Sun profile from observations, 
computing the contrast profiles, and assessing the effects of the center-to-limb 
variation (CLV) on the H$\alpha$ contrast profiles. Fundamentals of CM inversions 
are summarized in Sect.~\ref{SEC3}, which leads to the implementation of the inversion 
schemes in Sect.~\ref{SEC4}. We determine an optimized database of contrast profiles 
for CM inversions and introduce PCA for noise-stripping. Section~\ref{SEC5} presents 
physical maps determined from spectral line fitting and CM inversions, discusses them 
in the context of previous investigation \citep{Verma2012}, and and identifies systematic 
error sources, which may adversely affect the inversion results. In Sect.~\ref{SEC6}, our 
findings are placed in the context of previously published research, and we provide an 
outlook addressing spectral classification and database research.


\section{Observations}\label{SEC2}

The high-resolution spectra were obtained during the time period 2010 November 
18\,--\,23 with the VTT echelle spectrograph. An infrared grating with a 
51.6$^\circ$ blaze angle and 200 grooves mm$^{-1}$ was installed at the 
spectrograph. The spectra were recorded in the $12^\mathrm{th}$ order. Each 
spectrum consists of 2004 wavelength points, and with a dispersion of 6.0~m\AA\ 
pixel$^{-1}$, it covers a wavelength range of 12.0~\AA\ from 6559 to 6571~\AA. 
This spectral range includes two prominent absorption lines, i.e., H$\alpha$ and 
Fe\,\textsc{i} $\lambda$6569.2~\AA.

The H$\alpha$ spectra were recorded with a pco.4000 CCD camera. The full-format 
detector has 4008 $\times$ 2672 pixels with a size of 9~$\mu$m $\times$ 
9~$\mu$m. The spectra are digitized as 14-bit integers after applying 
2$\times$2-pixel binning. The resulting size of the spectra is 2004 $\times$ 
1168~pixels after some additional cropping in the spatial dimension. The image 
scale of the spectrograph is 8.99\arcsec\ mm$^{-1}$, and the step size for 
scanning a region-of-interest (ROI) in 244 steps is 0.32$\arcsec$, which yields a field-of-view (FOV) of 78\arcsec\ $\times$ 189\arcsec. Using a slit width of 
80~$\mu$m requires an exposure time of 300~ms for an appropriate utilization of 
the detector's full-well capacity. Ultimately, a cadence of 12~min can be 
achieved with this setup.

\begin{figure}[t]
\includegraphics[width=\columnwidth]{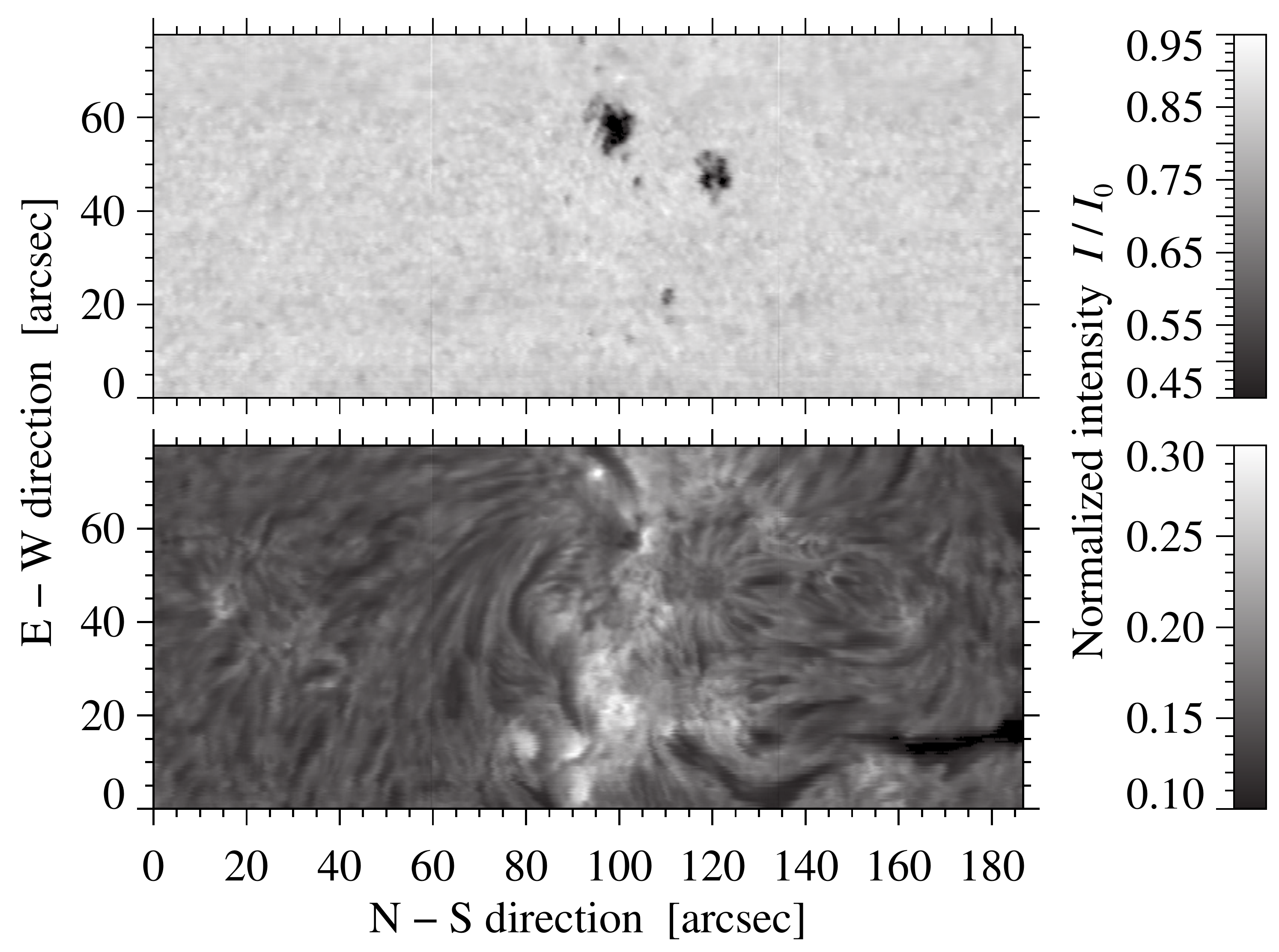}
\caption{Slit-reconstructed H$\alpha$ continuum (\textit{top}) and line-core 
    (\textit{bottom}) intensity maps of active region NOAA~11126 observed at
    10:23~UT on 2010 November~18. Two small decaying sunspots are visible in 
    the continuum intensity map, and an active-region filament is present in 
    bottom-right quadrant of the line-core intensity map. The cardinal
    directions do not follow the standard convention for solar images.}
\label{FIG01}    
\end{figure}

The decay of active region NOAA~11126 during the time period 2010 November 
18\,--\,23 was already studied by \citet{Verma2012} using Local Correlation 
Tracking (LCT) and spectral analysis. Echelle spectra were selected from this 
dataset, which were observed at 10:23~UT on 2010 November~18, to develop,
fine-tune, and evaluate our data processing pipeline. The observations 
targeted a sunspot group at heliographic coordinates (S$32.6^\circ$,
E$5.5^\circ$), where the cosine of the heliocentric angle is $\mu = 0.81$.
In the top and bottom panels of Fig.~\ref{FIG01}\ns, maps of
slit-reconstructed quasi-continuum and line-core intensity were compiled,
respectively. The decaying active region was accompanied by a filament,
partially visible in the lower-right corner of the H$\alpha$ line-core 
intensity map. The variety of features covered by the observations
makes this dataset an ideal choice for testing and evaluating all parts of 
the data processing pipeline.

\subsection{Dispersion and spectrograph profile}\label{SEC21}

The raw data of the VTT echelle spectrograph undergo standard data reduction 
steps including dark and flat-field corrections. Flat-field frames are captured 
while the pointing of the telescope is changing randomly around disk center. A 
few hundred dark and flat-field frames are averaged to minimize noise, and in 
the case of flat-field frames to reduce the impact of contrast features, which 
are still present in individual frames.

Preprocessing the echelle spectra involves continuum calibration and
correction, extraction of basic line properties, computing and calibrating 
the quiet-Sun profile, and conversion from intensity to contrast profiles. 
These steps are crucial to standardize the subsequent data processing and 
to facilitate a robust computation of physical maps and CM inversions.

Hair lines at the edges of the spectrum are used to correct the rotation of the 
spectrum. Position and width of the hair lines are determined with Gaussian 
fits. The resulting curves are smoothed before computing a linear fit and thus 
the rotation of the spectrum. The hair lines are removed from the spectrum by 
dividing by the respective Gaussian. Spectrograph tilt shears the individual 
spectra. Determining line-core position and width of (preferably) telluric lines 
with Gaussian fits allows us to shift the individual spectra by linear 
interpolation, thus effectively removing the spectrograph tilt. Division of the 
individual spectral profiles by the average spectral profile of the average 
flat-field frame yields the gain table, which is applied to all science spectra.

The wings of the H$\alpha$ line reach to about H$\alpha \pm 30$~\AA\ but only 
$\pm 6$~\AA\ are covered by the observed spectra. Depending on the level of 
solar activity, i.e., the presence of strong chromospheric Doppler signals, this 
wavelength range can be further reduced. A range of $\pm$3.8~\AA\ was deemed 
reasonable for an ROI containing small sunspots, pores, and an active region 
filament. Starting point to determine the spectral dispersion is the average 
flat-field spectrum, which is a good representation of the average quiet-Sun
profile $I_{\mathrm{qS},0}(\lambda)$ at disk center. A rough estimate can be 
gauged from the position of the spectral lines that were already used in the
correction of the spectrograph tilt. This is further refined by comparison 
with an atlas spectrum of a Fourier transform spectrometer \citep{Wallace1998}. 
The linear correlation between atlas and observed spectrum is calculated while
adjusting the initial estimate for the dispersion in small steps. The highest
correlation yields an improved value for the dispersion.

Broad-band interference filters were used instead of the predisperser masks 
to avoid overlapping spectral orders. Thus, the full length of the slit can be 
exploited to record spectral scans. Therefore, the task is to correct the spectra
for the combined spectrograph and interference filter profiles. The ratio 
of observed and atlas spectra is first smoothed with a Gaussian kernel and 
secondly, the envelope is computed to eliminate large variations. Further 
iterative boxcar smoothing produces an even correction curve that matches the 
observed quiet-Sun spectrum at disk center to the atlas spectrum.


\subsection{Preprocessing of spatio-spectral data}\label{SEC22}

Preprocessing of the spatio-spectral data cube included so far dark
correction, intensity calibration by means of the gain table, spectrum rotation,
and removal of the spectrograph tilt. The ``pseudo''-continuum is derived for 
the blue and red wings, taking into account that the H$\alpha$ line is too broad
to determine the actual continuum intensity. Thus, two small wavelength ranges 
are selected in the wings, which are free of any solar or telluric line. The two
pseudo-continuum positions make it possible to remove an intensity 
gradient from the spectra. Furthermore, the spectra are normalized 
such that the local continuum refers to unity. This step also includes a 
correction for variations of the sky brightness. The normalization coefficients 
are kept in a two-dimensional map of the continuum intensity. Henceforth, we can 
determine from the minimum of the H$\alpha$ line the line-core intensity and 
position, i.e., a proxy line-core Doppler velocity with an accuracy given 
by the wavelength sampling.


\subsection{Quiet-Sun profile and stray light correction}\label{SEC23}

In general, science data are not taken at disk center, i.e., the previously 
determined quiet-Sun profile is not applicable due to the CLV of spectral line
profiles. Thus, quiet-Sun regions have to be identified in the FOV for computing
an average quiet-Sun profile suitable for the specific location on the solar 
disk. The requirements are that continuum intensity is between the 
$20^\mathrm{th}$ and $95^\mathrm{th}$ percentile, the line-core intensity is
between the $30^\mathrm{th}$ and $90^\mathrm{th}$ percentile, and the proxy
line-core Doppler velocity is between the $20^\mathrm{th}$ and $80^\mathrm{th}$
percentile. This ensures that sunspots, pores, bright points, faculae, filigree,
dark mottles, any type of filament, and high-velocity features are excluded from 
computing the average quiet-Sun profile. The Doppler shift of individual 
profiles is corrected before taking the average. This procedure was tested 
by us using several other datasets, and the underlying assumptions are
reasonable for a wide variety scenes on the solar surface. The 
percentiles have to be adjusted if bright/dark intensity or high-velocity 
features dominate the FOV. 

\citet{David1961} tabulated the CLV of quiet-Sun H$\alpha$ spectra 
for seven values of $\mu$. In principle, the profile for $\mu = 0.8$
can be used for the present data. However, an interpolation step was added
in the data processing pipeline to generalize this approach for datasets that 
do not match the tabulated $\mu$ values. The interpolated quiet-Sun
profile was used to match continuum and line-core intensity of the observed 
profile. In addition, the width of both profiles were matched. This corresponds
to an approximate correction of the spectrograph modulation transfer function 
and of the instrument's stray light contribution. The choice of a proper
background profile is crucial in the subsequent CM inversions. In preparation
for CM inversion, the contrast profiles were computed according to
\begin{equation}
C(\lambda) = \frac{I(\lambda)-I_\mathrm{qS}(\lambda)}{I_\mathrm{qS}(\lambda)},
\label{EQN1}
\end{equation}
where $I(\lambda)$ and $I_\mathrm{qS}(\lambda)$ are the observed and quiet-Sun 
intensity profiles, respectively. Finally, two-dimensional maps of physical 
properties are (re)computed including continuum intensity, line-core intensity, 
equivalent width, absolute contrast, and a proxy for the Doppler velocity.

The \textit{sTools} software library \citep{Kuckein2017} was originally 
developed for the data pipeline of the GREGOR Fabry-P\'erot Interferometer 
\citep[GFPI,][]{Denker2010, Puschmann2012} and the High-resolution Fast Imager 
\citep[HiFI,][]{Denker2018}. Nowadays, almost all solar software development 
at AIP is integrated into this software library, including reduction and analysis 
of VTT echelle spectra.


\subsection{Center-to-limb variation of quiet-Sun profiles}\label{SEC24}

The CLV of quiet-Sun H$\alpha$ profiles is a central issue for CM inversions. 
The CLV can be derived either from observations or from radiation hydrodynamic 
models and radiative transfer calculations, e.g., many codes and models are 
available for cool-stars atmosphere synthesis such as the simple FAL-C model 
\citep{Vernazza1981} or the state-of-the-art BIFROST numerical simulations 
\citep{Gudiksen2011}.

In this study, we use the H$\alpha$ background irradiation from observations 
based on the work of \citet{David1961}. These profiles were obtained in years
1958\,--\,1959 together with higher hydrogen Balmer lines 
H$\beta$\,--\,H$\delta$ with the G\"{o}ttingen Solar Tower
\citep{tenBruggecate1958} equipped with the 8-meter Littrow spectrograph
\citep{vonAlvensleben1957}. The H$\alpha$ profiles are tabulated 
on a non-equidistant grid of 55 wavelengths points in a range of 0\,--\,30~\AA\ 
relative to the line center.

\begin{figure}[t]
\includegraphics[width=\columnwidth]{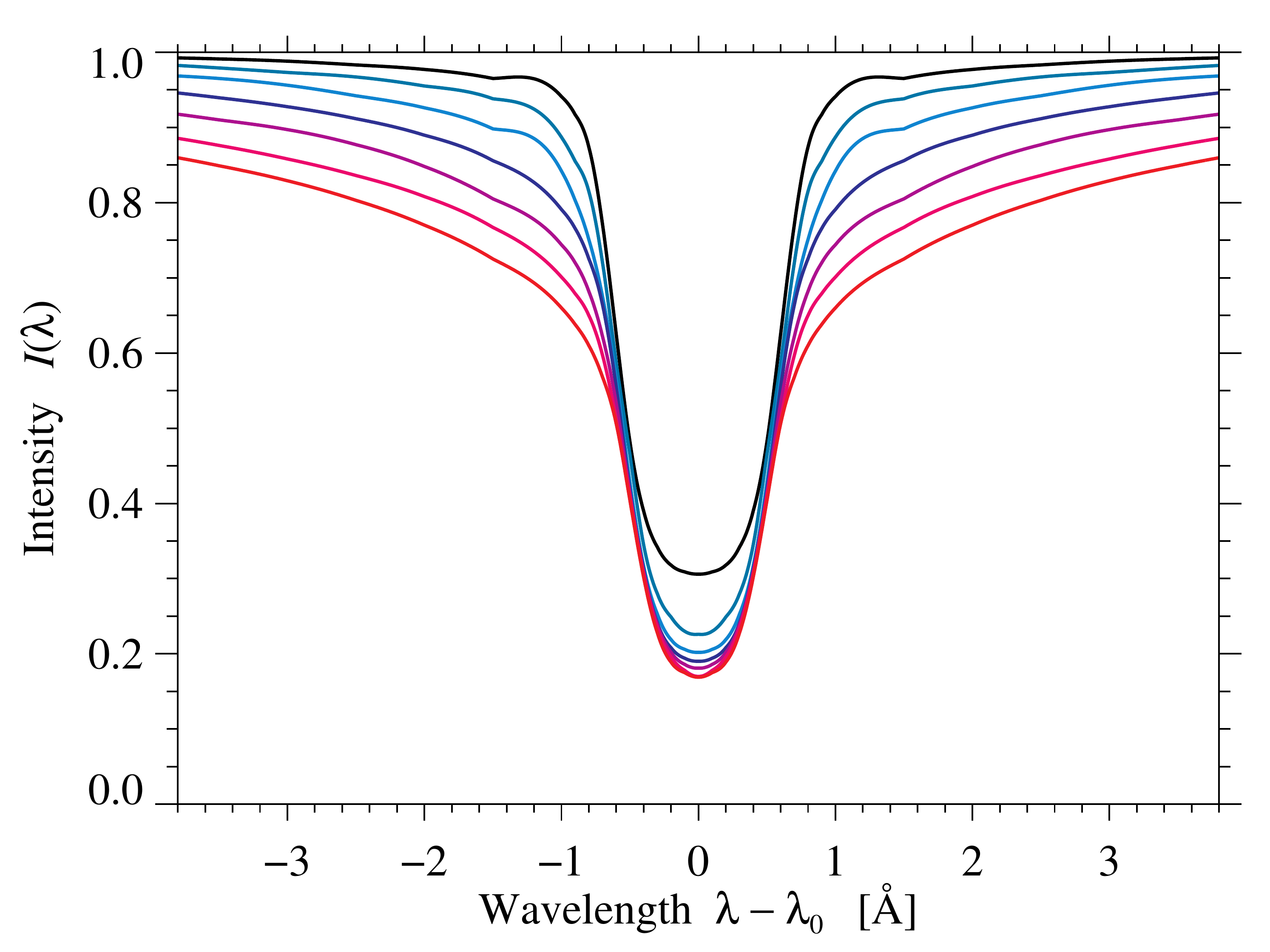}
\caption{Center-to-limb variation of H$\alpha$ absorption profiles in the  
    spectral range H$\alpha \pm 3.8$~\AA. Colors correspond to the seven positions
    from disk-center (\textit{red}) to near the limb (\textit{black}), where the
    cosine of the heliocentric angle is $\mu = $ 1.0, 0.8, 0.6, 0.436, 0.312, 
    0.28, and 0.141.}
\label{FIG02}    
\end{figure}

The profiles are tabulated in Table~1 of \citet{David1961} for seven heliocentric angles. 
The cosines of these seven angles are $\mu =$ 1.0, 0.8, 0.6, 0.436, 0.312, 0.28, 
and 0.141. The CLV of the inner part of the H$\alpha$ line profile is given in 
Fig.~\ref{FIG02}\ns, where the deepest profile in red corresponds to disk-center 
observations. The CLV clearly affects line-core intensity and the shape of the 
line wings. According to various studies the line core is formed at approximately 
1700~km above the solar surface \citep{Leenaarts2012, Vernazza1981, White1966}. 
There is neither a sharp boundary nor an exact height because of the dynamic nature 
of the chromosphere and magnetic activity. In principle, radiative transfer in the 
chromosphere is by nature three-dimensional.

\section{Cloud Model inversions}\label{SEC3}

The classic CM was proposed by \citet{Beckers1964} as a convenient tool for 
determining the physical properties of cloud-like structures of absorbing 
material suspended by the magnetic field above the solar surface. Albeit highly 
simplified, this model is still commonly used and delivers physical insight into 
dynamic processes in the solar chromosphere. The CM is based on the following 
important assumptions, i.e., the plasma cloud is located above and is fully 
separated from the chromospheric forest \citep{Tziotziou2007, Chae2014}, the 
plasma parameters are constant along the LOS, and the incident irradiation 
originating from underneath the cloud has the same properties as the radiation 
from the surrounding undisturbed atmosphere. The last condition highlights
the importance of the selecting a suitable quiet-Sun background profile
\citep[cf.,][]{Bostanci2010}. Finally, the four parameters defining the 
radiative transfer and line formation in the CM are the optical thickness 
$\tau_0$, Doppler velocity of the cloud $v_\mathrm{D}$, Doppler width 
$\Delta\lambda_\mathrm{D}$ of the absorption profile, and source function $S$.

\begin{figure*}[t]
\includegraphics[width=\textwidth]{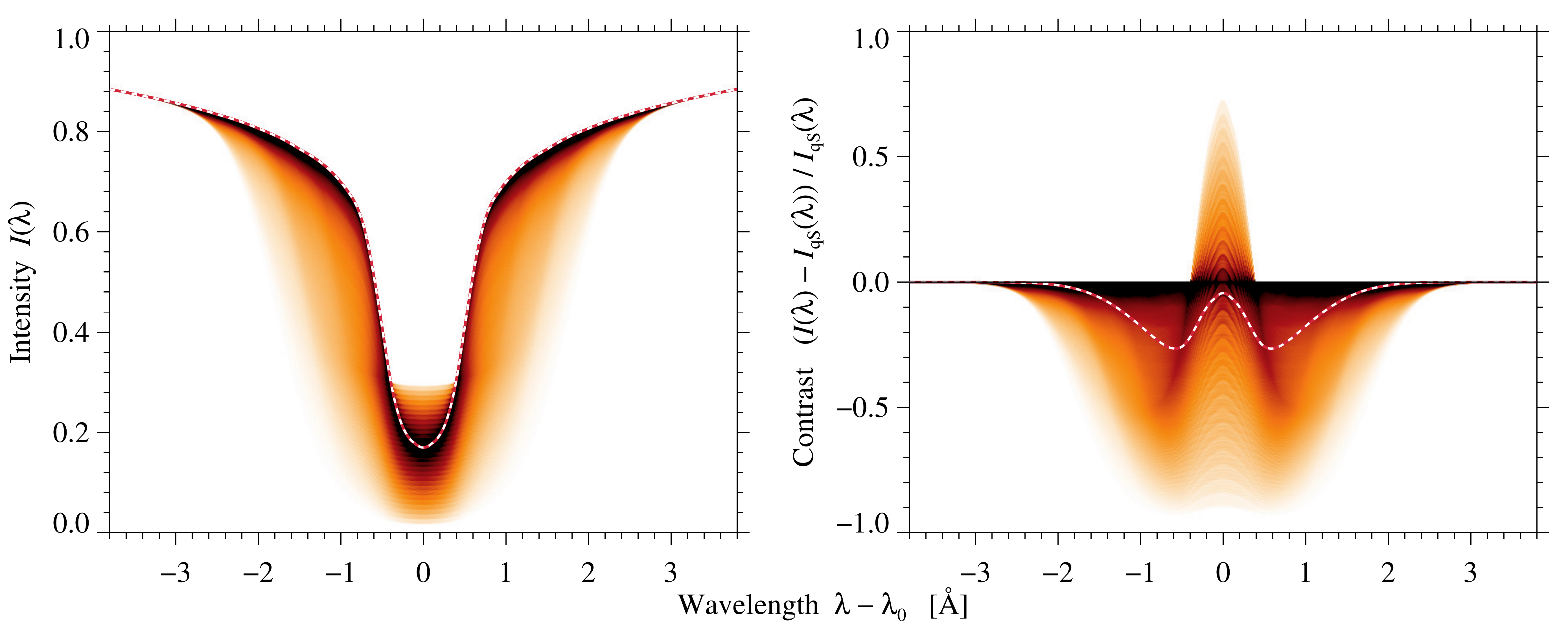}
\caption{Bulk density distributions of H$\alpha$ intensity (\textit{left})
    and contrast (\textit{right}) profiles, which summarize the optimized CM
    database. The two-dimensional histograms display the diversity of 
    shapes and features that the CM can reproduce. The red-white dashed 
    curves refer to quiet-Sun intensity and average contrast profiles,
    respectively.}
\label{FIG03}    
\end{figure*}

Some characteristics of strong chromospheric absorption lines are easier to 
discover in contrast profiles $C(\lambda)$ (Eqn.~\ref{EQN1}). Note that the 
contrast profile is defined independently of the CM, which is a useful property 
in noise-stripping of spectral profiles using PCA. The CM provides a 
relationship between the contrast profile and the four free fit parameters:
\begin{eqnarray}
C(\lambda)    & = & \left[ \frac{S}{I_\mathrm{qS}(\lambda)} - 1 \right]
                    \Big( 1 -\exp[-\tau(\lambda)] \Big)
                    \quad {\rm with} \label{EQN_CM} \\
\tau(\lambda) & = & \tau_0 \exp \left[ - \left(
                    \frac{\lambda-\lambda_\mathrm{D}}{\Delta\lambda_\mathrm{D}} \right)^2 \right]
\label{EQN2}
\end{eqnarray}
The LOS velocity $v_\mathrm{D}$ of the cloud can be derived, once 
$\lambda_\mathrm{D}$ is known, according to
\begin{equation}
v_\mathrm{D} = c\, \frac{\lambda_\mathrm{D} - \lambda_0}{\lambda_0},
\label{EQN3}
\end{equation}
where $\lambda_0$ is the central wavelength of the strong chromospheric
absorption line and $c$ the speed of light in vacuum. In summary, 
the absorption profiles deliver information about light-matter interactions in 
the solar plasma. As Eqns.~\ref{EQN1}\,--\,\ref{EQN3} demonstrate, the 
contrast profile formulation represents a simplified version of the radiative 
transfer equation, taking into account the aforementioned model simplifications.
These equations are contrived in terms of the observed intensity $I(\lambda)$ 
of the dynamic cloud-like feature and the undisturbed quiet-Sun background 
reference frame $I_\mathrm{qS}(\lambda)$. A simple CM inversion approach is 
non-linear least-squares fitting, for example, with the MPFIT software package 
\citep{Markwardt2009}, where the quiet-Sun background profile is handed to
the fitting routine in a structure as private data 
\citep[see e.g.,][]{GonzalezManrique2017}. However, the iterative nature of 
the algorithm and the tendency to get lost in local extrema render this 
procedure inefficient. The resulting computation times may be acceptable 
for individual datasets but not for bulk processing of billions of spectral 
profiles.


\section{Implementation}\label{SEC4}


\subsection{Creating the Cloud Model database}\label{SEC41}

Starting point for creating the CM database is the desired wavelength range 
(H$\alpha \pm 3.8$~\AA) and the quiet-Sun profile that matches the cosine of the 
heliocentric angle of the observations ($\mu = 0.81$ in the present case) as described at the 
beginning of Sect.~\ref{SEC2} and in Sect~\ref{SEC24}. In addition, a coarser 
spectral sampling reduces significantly the computation time. In most cases, a 
sampling of 10~m\AA\ pixel$^{-1}$ is sufficient. Consequently, observations at 
different locations on the solar surface or with other observing parameters 
require a different CM database.

The four CM parameters were sampled on an equidistant grid in the following 
intervals: (1) $\tau_0 \in [0.1,\ 3.0]$ with an increment $\delta\tau_0 = 0.1$, 
(2) $v_\mathrm{D} \in [-40.0,\ +40.0]$~km s$^\mathrm{-1}$ with an increment $\delta 
v_\mathrm{D} = 4.0$~km s$^\mathrm{-1}$, (3) $\Delta\lambda_\mathrm{D} \in [0.2,\ 
1.0]$~\AA\ with an increment $\delta\Delta\lambda_\mathrm{D} = 0.04$~\AA, and (4) 
$S \in [0.01,\ 0.3]$ with an increment $\delta S = 0.01$. This selection of the 
CM parameters is in agreement with many other CM studies. The upper limits of 
the parameter space may already border at the upper limit for physically 
reasonable values, with the exception of the Doppler velocity.

Initially, both contrast and intensity profiles are computed for all possible 
combinations of CM parameters, thus producing a large set of $2 \times 396\,900$ 
profiles. Not all combinations are plausible or carry enough significance, and 
their presence will only obstruct fast data processing and upcoming CM 
inversions. The goal is to optimize the database with the help of PCA. The first 
ten eigenvectors were derived with PCA for all contrast profiles contained in the 
database. The PCA was carried out on the sum-of-squares (SSQ) and the cross-products 
matrix. Each contrast profile was restored using the projections of the profiles 
onto the first ten principal components. The SSQ was computed for the 
difference of the restored contrast profiles and those in the database. Only 
80\% of the profiles with lowest deviation were kept. The linear correlation 
serves as a second selection criterion, which discards profiles, where the 
correlation coefficient is below 0.8. The reduced database is again subjected to 
PCA and another optimization cycle. This time with limits of 80\% for the 
SSQ deviations and 0.95 for the linear correlation. The thresholds for the linear
correlation coefficients were validated by visual inspection of the large number of 
fitted contrast profiles. The remaining $2 \times 253\,038$ contrast and intensity profiles 
form the final version of the database. The first 10 eigenvectors representing the contrast
profiles of the optimized CM database are kept for CM inversions and noise-stripping.

\begin{figure*}
\includegraphics[width=\textwidth]{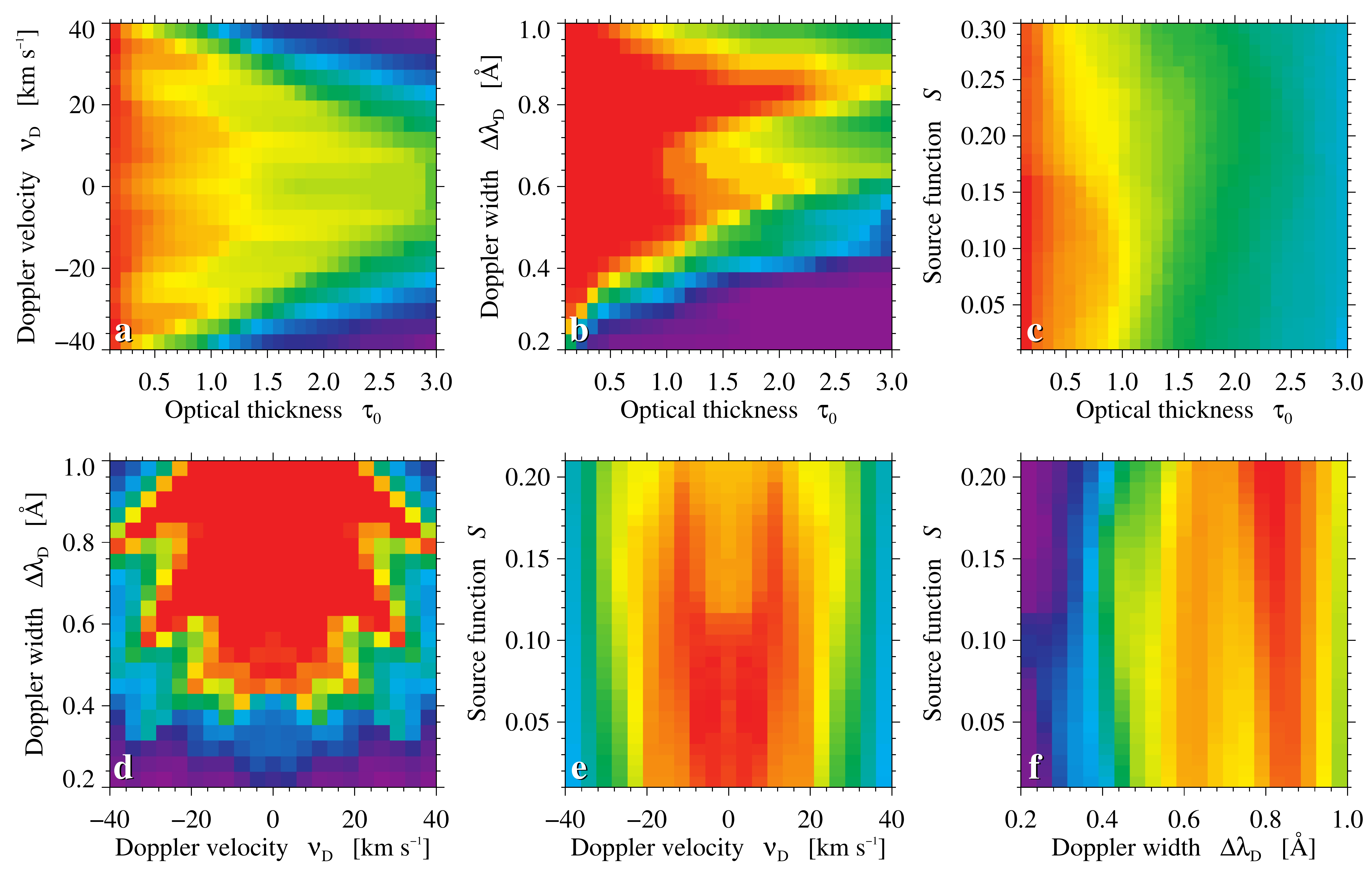}
\caption{Two-dimensional histograms displaying projections of the 
    four-dimensional CM parameter space. The six possible projections are: (a)
    Doppler velocity $v_\mathrm{D}$ vs.\ optical thickness $\tau_0$, (b) Doppler
    width $\Delta\lambda_\mathrm{D}$ vs.\ optical thickness $\tau_0$, (c) source
    function $S$ vs.\ optical thickness $\tau_0$, (d) Doppler width
    $\Delta\lambda_\mathrm{D}$ vs.\ Doppler velocity $v_\mathrm{D}$, (e) source
    function $S$ vs.\ Doppler velocity $v_\mathrm{D}$, and (f) source function 
    $S$ vs.\ Doppler width $\Delta\lambda_\mathrm{D}$.}
\label{FIG04}
\end{figure*}

Figure~\ref{FIG03}{\ns} summarizes the optimized database in form of two-dimensional 
histograms, where each profile contributes with 761 spectral points across the 
depicted wavelength range. The color scale from black over dark to light orange 
indicates the frequency of occurrence from high to low. The majority of profiles 
contained in the database stays close to the quiet-Sun background profile (red-white dashed 
curve in the left panel of Fig.~\ref{FIG03}\ns). The distribution of the 
intensity profiles shows that the database contains narrow as well as broader 
profiles. Since the Doppler velocities are equally distributed around zero, the 
superposition of red- and blue-shifted profiles will yield a symmetric 
two-dimensional histogram.

Any profile that is not contained within the black-orange area cannot be derived 
with CM inversions. In principle, a wider range of Doppler velocities enlarges 
the black-orange area. However, the extra contrast and intensity profiles are 
inevitably related to eruptive events. In any case, moustache profiles with 
enhanced line-wing emission or profiles with emission cores are outside the 
reach of CM inversions. Evidently, if the quiet-Sun profile is too broad, 
narrower observed profiles cannot be fitted. More importantly, a strong bias 
will be introduced in all four CM parameters.

The distribution of contrast profiles in the right panel of Fig.~\ref{FIG03}{\ns} 
is more complex and dominated by \textsf{W}-shaped profiles. As the quiet-Sun 
contrast profile will be identical to zero, only the average contrast profile is 
plotted as a red-white dashed curve. Positive contrasts are rare and only occur 
near the central wavelength. They typically result from asymmetric contrast 
profiles with moderate red- and blue-shifts (below $\pm$10~km s$^{-1}$).

Figure~\ref{FIG04}{\ns} shows two-dimensional histograms of the six possible 
projections of the four-dimensional CM parameter space, which make it possible to 
evaluate the iterative optimization of the database. The frequency of occurrence is 
given by rainbow colors, where red and violet indicate high and low number 
densities, respectively. Sparser regions in the CM parameter space are 
characterized by high velocities and large optical thickness and by low Doppler 
width and large optical thickness. The relationship between Doppler width and 
velocity is more complex and characterized by ridges with high number density. 
The source function shows only a weak dependence of the number density with 
the remaining three CM parameters. In general, profiles with a Doppler width  
$\Delta\lambda_\mathrm{D} < 0.4$~\AA\ and with an optical thickness above 
$\tau_0 > 2.0$ are rare.

\begin{figure}[t]
\includegraphics[width=\columnwidth]{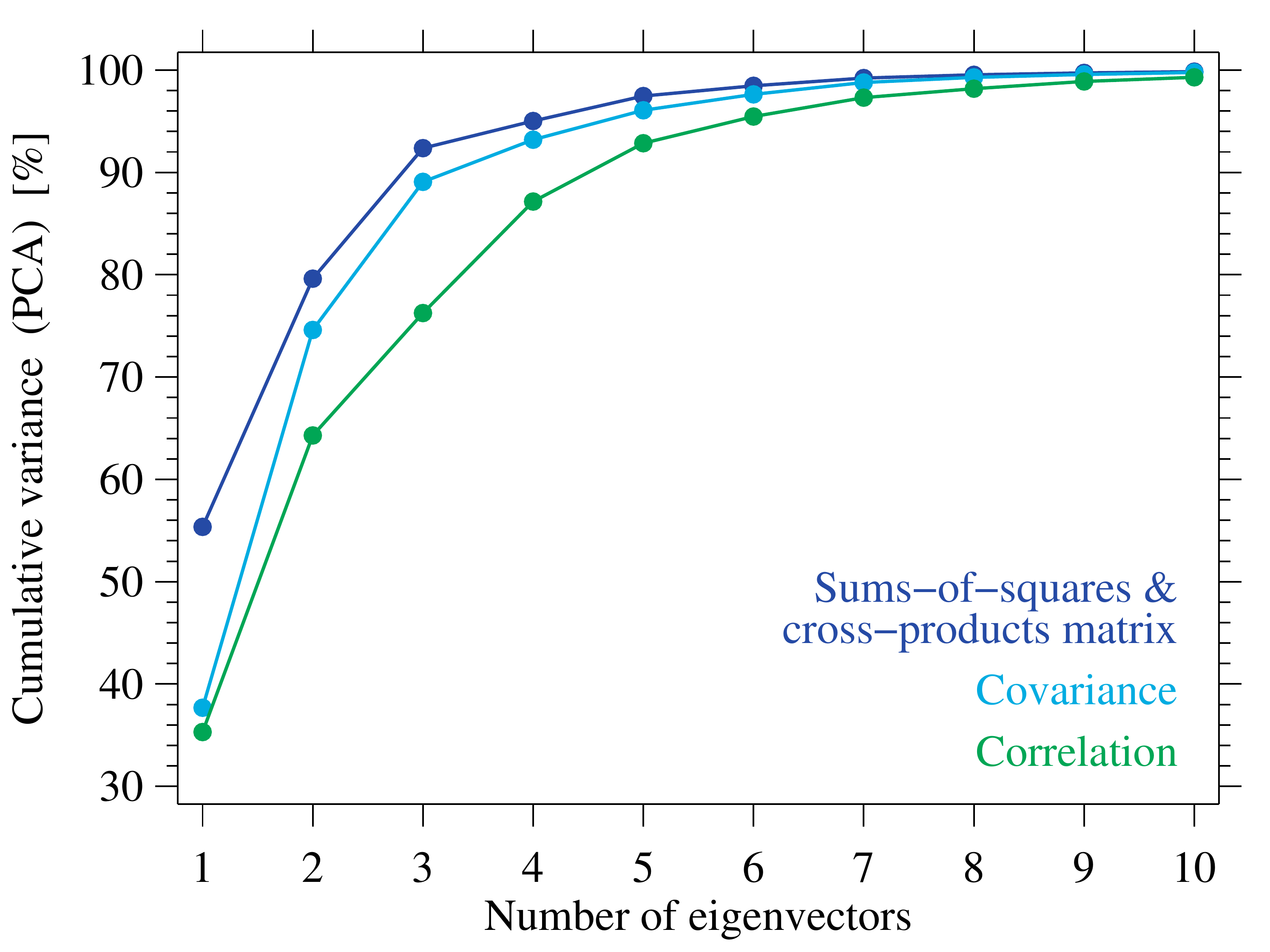}
\caption{Percentage of the variation in the CM database of contrast
    profiles expressed by the first ten eigenvectors for three different 
    PCA options.}
\label{FIG05}
\end{figure}


\subsection{Assessment of Principal Component Analyses}\label{SEC42}

The choice of contrast profiles as input for PCA decomposition is
motivated by the fact that division by the quiet-Sun background profile in 
Eqn.~\ref{EQN1} significantly reduces undesirable signatures of telluric 
and photospheric spectral lines. By default PCA is carried out on the 
covariance matrix. Less commonly used 
options include the correlation matrix and the SSQ and cross-products matrix. 
The performance of these three options was evaluated in a numerical experiment.
Figure~\ref{FIG05}{\ns} is a visual representation of how much of the
variance is explained by an increasing number of eigenvectors. Variance in 
this context refers to the diversity of contrast profiles, which is contained 
in the optimized CM database. This is plotted for the first ten eigenvectors,
which clearly favors the SSQ option. Already the first SSQ eigenvector
exceeds the variance of the other two options by more than 20 percentage 
points. The SSQ curve for the cumulative variance is always above the other
two curves. All curves start to converge after including five eigenvectors 
and are virtually the same for ten or more eigenvectors.

\begin{figure*}[t]
\includegraphics[width=\textwidth]{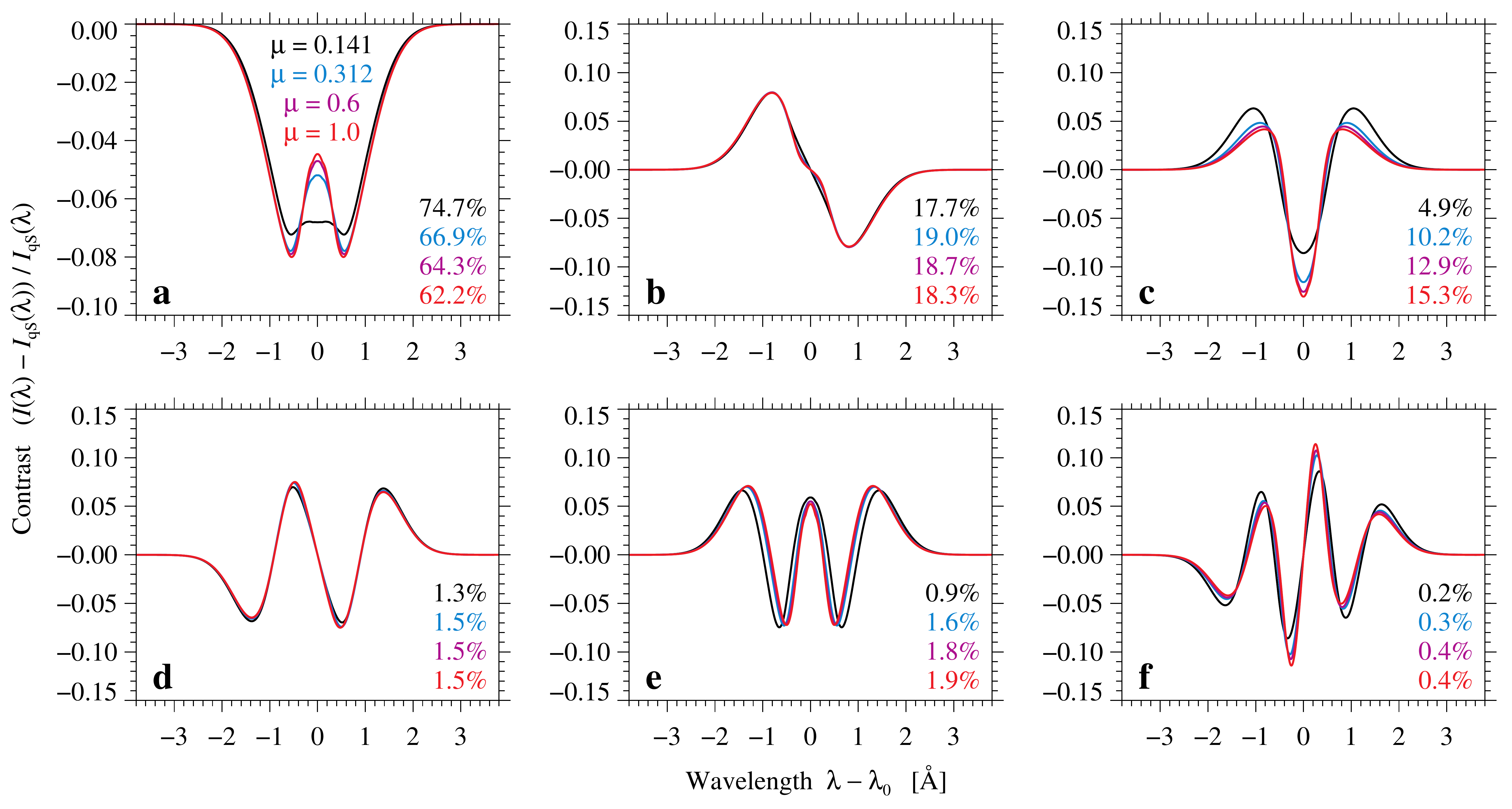}
\caption{Center-to-limb variations of the first six eigenvectors for a 
    selection of four $\mu$ values. The color code corresponds to
    Fig.~\ref{FIG02}\ns, and the $\mu$ values are given in the upper-left 
    panel inside the eigenvectors. Note that in some cases the sign 
    of the eigenvector was switched and the order of the eigenvectors was 
    sometimes changed, which improves the legibility of display.The
    variance of the contrast profiles for the four $\mu$ values that can be
    explained by the eigenvectors is given in the lower-right corner of each
    panel.}
\label{FIG06}
\end{figure*}

The eigenvectors are typically arranged in descending order according to their 
ability to express the variance in the CM database. Motivated by the 
results displayed in Fig.~\ref{FIG05}\ns, the first ten eigenvectors are 
sufficient to reconstruct profiles contained in the database. As a
reference, the first six eigenvectors and their CLV are presented in 
Fig.~\ref{FIG06}{\ns}. The first eigenvector has a distinct \textsf{W}-shape,
whereas the other eigenvectors display an increasing number of maxima and 
minima. Thus, only the first six eigenvectors are depicted because the 
four left-out eigenvectors have the same pattern. The sign of the eigenvectors 
is not important so that the sign was switched whenever necessary for 
clarity. 

In Sect.~\ref{SEC24}, emphasis was placed on the CLV of the quiet-Sun background 
profile, which prompted another numerical experiment. Four quiet-Sun background 
profiles ($\mu = 1.0$, 0.6, 0.312, and 0.141) were selected from the work 
of \citet{David1961}. Individual databases were derived for this selection, and 
four sets of eigenvectors were computed. The first six eigenvectors for 
the four different $\mu$ values are also shown in Fig.~\ref{FIG06}\ns. 

In general, the eigenvectors are zero outside the interval $[-3.0,\, 
+3.0]$~\AA. The first set of eigenvectors resembles the average contrast profile 
in the right panel of Fig.~\ref{FIG03}\ns. The major differences in 
Fig.~\ref{FIG06}{\ns}a are related to the depth of the central local maximum 
and of the neighboring minima. This is not surprising because the core of the 
H$\alpha$ line showed the largest variation from disk center to the limb (see 
Fig.~\ref{FIG02}\ns). The asymmetric eigenvectors in Fig.~\ref{FIG06}{\ns}b are 
virtually identical. They are mainly responsible for modifying the restored 
contrast profiles according to the Doppler shift. Notable changes among 
the eigenvectors appear in Fig.~\ref{FIG06}{\ns}c, where the core contrast is 
affected and also the neighboring local maxima. The next asymmetric 
eigenvectors in Fig.~\ref{FIG06}{\ns}d show a similar behavior as the second set 
of eigenvectors. In comparison with the first set of eigenvectors, the 
\textsf{W}-shaped eigenvectors in Fig.~\ref{FIG06}{\ns}e exhibit positive and 
negative contrasts. In addition, the eigenvector closest to the limb is 
slightly broader. Finally, the asymmetric eigenvectors in
Fig.~\ref{FIG06}{\ns}f start to show a larger variation than their 
simpler counterparts. The variance of the contrast profiles that can be
explained by the eigenvectors is given for each eigenvector and $\mu$ value. 
Close to the limb, already the first two eigenvectors explain 92.4\% of the
variance, whereas the second and third eigenvector become almost the same near
disk center. The pattern established by the first six sets of eigenvectors
continues, and the number of extreme values increases. In summary, the CLV 
of eigenvectors displays a significant variation, which will lead to 
noticeable errors in the CM inversions, if not taken properly into 
account. 

\subsection{Derivation of Cloud Model parameters}\label{SEC43}

The last part of the data processing pipeline concerns the CM inversions. The CM database includes H$\alpha$ intensity and contrast profiles as well as the corresponding CM parameters. If only the observed spectra serve as input, a first iteration of the Levenberg-Marquardt technique \citep{Markwardt2009} is used to solve the least-squares problem (Sect.~\ref{SEC3}). The initial estimates for the CM parameters are $\tau_0 = 1.5$, $\Delta\lambda_\mathrm{D} = 0.6$~\AA, $S = 0.15$, and $v_\mathrm{D}$ is taken from line-core fitting. In addition, the CM parameters were restricted to the intervals $\tau_0 \in [0.1,\ 4.0]$, $v_\mathrm{D} \in [-50.0,\ +50.0]$~km s$^\mathrm{-1}$, $\Delta\lambda_\mathrm{D} \in [0.1,\  1.5]$~\AA, $S \in [0.01,\ 0.5]$. Fits were rejected if the CM parameters converged at the limits of the respective interval, if the linear correlation coefficient between observed and fitted contrast profile was $r < 0.95$, or if the rank-order correlation coefficient was $\rho < 0.4$. In the second iteration, the initial estimates are improved assuming linear relationships between $v_\mathrm{D}$ and line-core Doppler velocities, $\Delta\lambda_\mathrm{D}$ and the FWHM of the observed intensity profiles, and $\tau_0$ and the equivalent contrast of the contrast profile. The equivalent contrast is defined in analogy to the equivalent width of a spectral line, i.e., by computing the area under the contrast profile and expressing it as a rectangle with unit height and a width in {\AA}ngstr\"om. The estimate for the source function $S$ is simply the median of the observed distribution. Good fits were identified again according to the criteria that were applied after the first iteration. This simplified CM inversion scheme already yields reasonably good results. In addition, the computations are relatively fast because the Levenberg-Marquardt fits are carried out only twice per contrast profile. All other computations are short compared to this part of the inversions.

The CM database and PCA decomposition are the ingredients for a further refinement of CM inversions. The entries in the CM database are sorted according to the equivalent contrast of the contrast profiles, and the ten coefficients for the PCA are added as additional entries into the database. The equivalent contrast robustly shrinks the range of contrast profiles in the CM database that corresponds to the observed profiles. Typically, 1000 profiles are selected, for which the ten best matches are identified with good convergence of the ten PCA coefficients. Conformity is based on the smallest SSQ, which does not favor strong outliers of the  PCA coefficients. In principle, selecting the best match can be considered as rough CM inversion with an accuracy corresponding to the grid spacing of the CM database (see Sect.~\ref{SEC41}).

Since ten Levenberg-Marquardt fits are required per contrast profile to find the best match, this method is five times slower than the more simplistic approach mentioned above. However, the combination of CM database and PCA decomposition avoids getting trapped in local minima. Good fits were identified again according to the criteria mentioned above. All attempts to improve the inversion results, e.g., using CM parameters from a local neighborhood, produced ``improvements'' only at the level of numerical noise. 

\begin{figure*}[t]
\includegraphics[width=\textwidth]{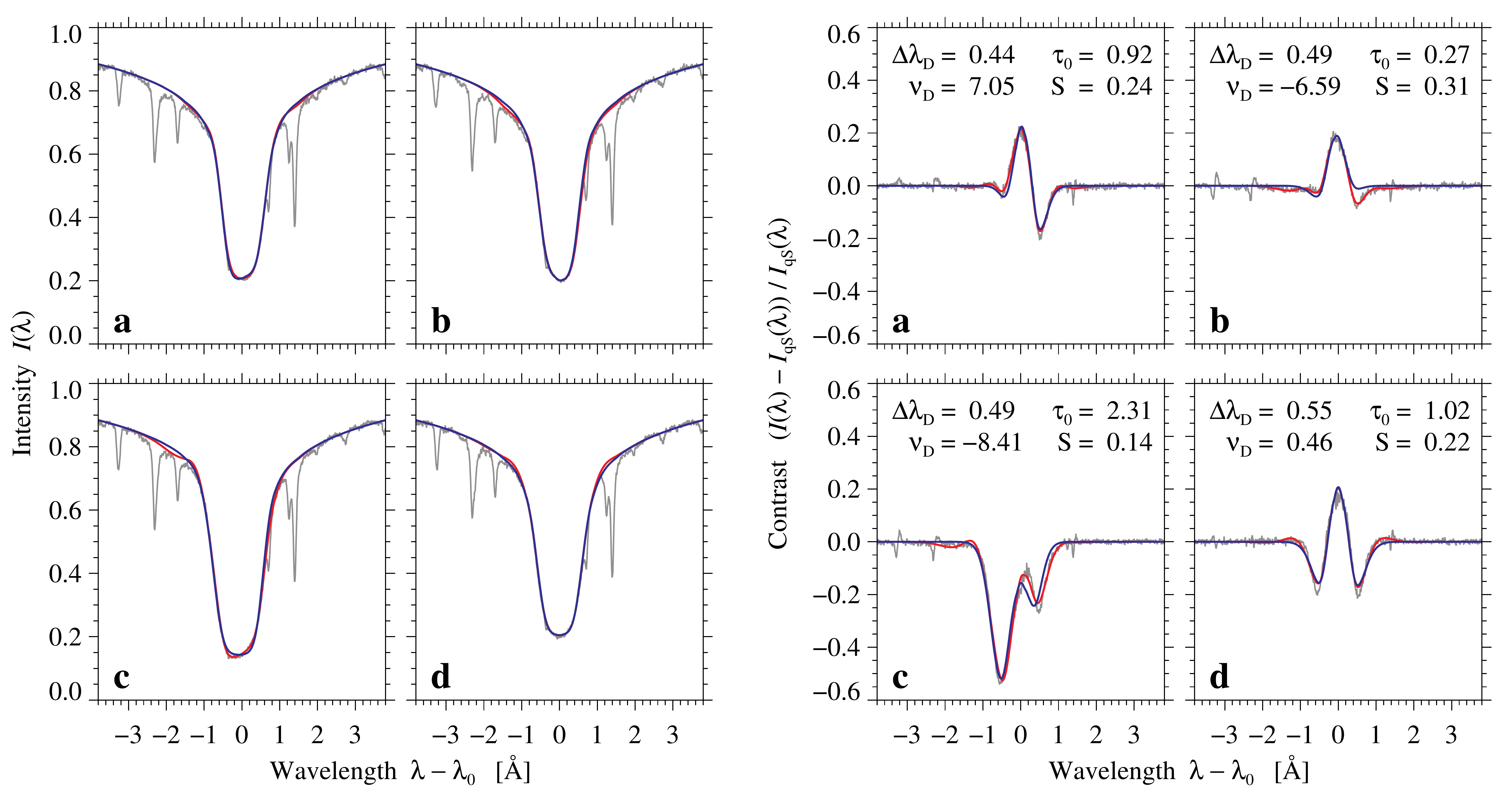}
\caption{Four observed H$\alpha$ intensity profiles (\textit{left}) along
    with the corresponding contrast profiles (\textit{right}). The observed 
    spectra (\textit{gray}) contain blends of telluric and solar spectral 
    lines. After PCA decomposition the contrast profiles are well represented 
    (\textit{red}) and the noise was eliminated. he CM inversions 
    (\textit{blue}) yielded the four CM parameters, which are displayed in
    the panels of the contrast profiles. The units correspond to those 
    specified in Sec.~\ref{SEC41}. The H$\alpha$ intensity profiles were 
    restored according to Eqn.~\ref{EQN1} with the appropriate quiet-Sun 
    H$\alpha$ background profile \citep{David1961}.}
\label{FIG07}
\end{figure*}

Compared to \citet{GonzalezManrique2017}, the current implementation of the CM inversions includes a refinement step that links spectral line to CM parameters. Furthermore, the CM database is organized according to the absolute contrast so that only a subset of the CM contrast profiles has to be compared to the observed ones. Finally, the current inversion scheme incorporates noise-stripping based on PCA.


\section{Results}\label{SEC5}

The estimated noise level of the H$\alpha$ intensity profiles is $7.5 \times 10^{-3}$ of the quiet-Sun continuum intensity. Noise-stripping is performed by computing the coefficients for the eigenvectors using Singular Value Decomposition (SVD) to solve the set of simultaneous linear equations between the contrast profile on one side and the eigenvectors on the other side of the equation. The fitted contrast profiles are just the sum of the ten eigenvectors weighted by the coefficients. The H$\alpha$ intensity profiles can be restored according to Eqn.~\ref{EQN1} using the fitted contrast profile and the appropriate quiet-Sun background profile. Since the observed quiet-Sun profile still contains noise and is contaminated by blends of telluric and solar spectral lines, either the quiet-Sun profile must be smoothed and filtered or one directly resorts to a suitable quiet-Sun background profile from \citet{David1961}. Thus, a noise-free spatio-spectral data cube is now available for CM inversions, besides the unaltered preprocessed data cube. Comparing the CM results of these two spatio-spectral data cubes enables estimates of systematic errors inherent to the data processing procedure.

\begin{figure}[t]
\includegraphics[width=\columnwidth]{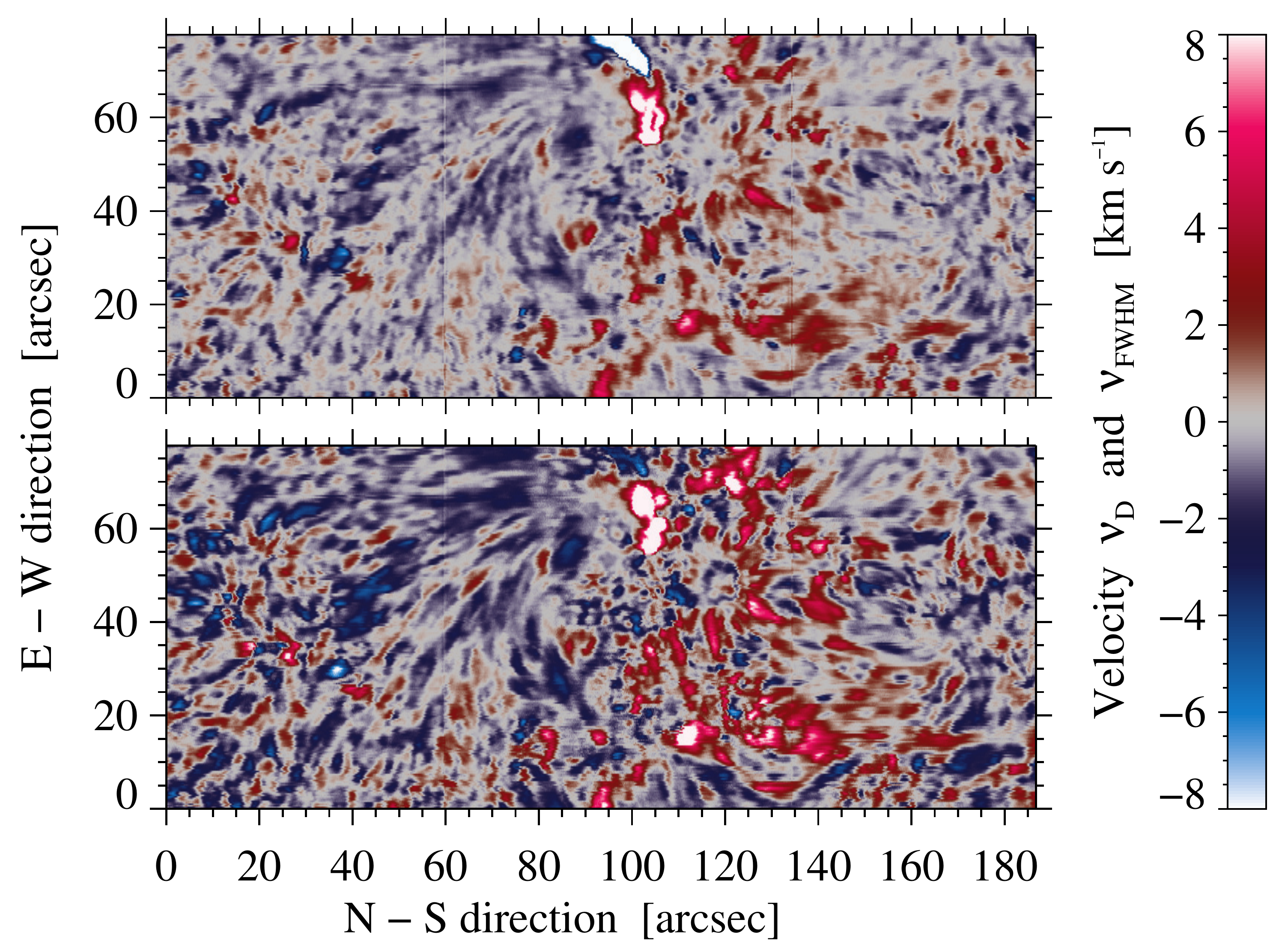}
\caption{Two-dimensional H$\alpha$ LOS velocity maps computed from line-core
    fits (\textit{bottom}) and from bisector analysis at the FWHM level 
    (\textit{top}).}
\label{FIG08}
\end{figure}

\begin{figure*}[t]
\includegraphics[width=\textwidth]{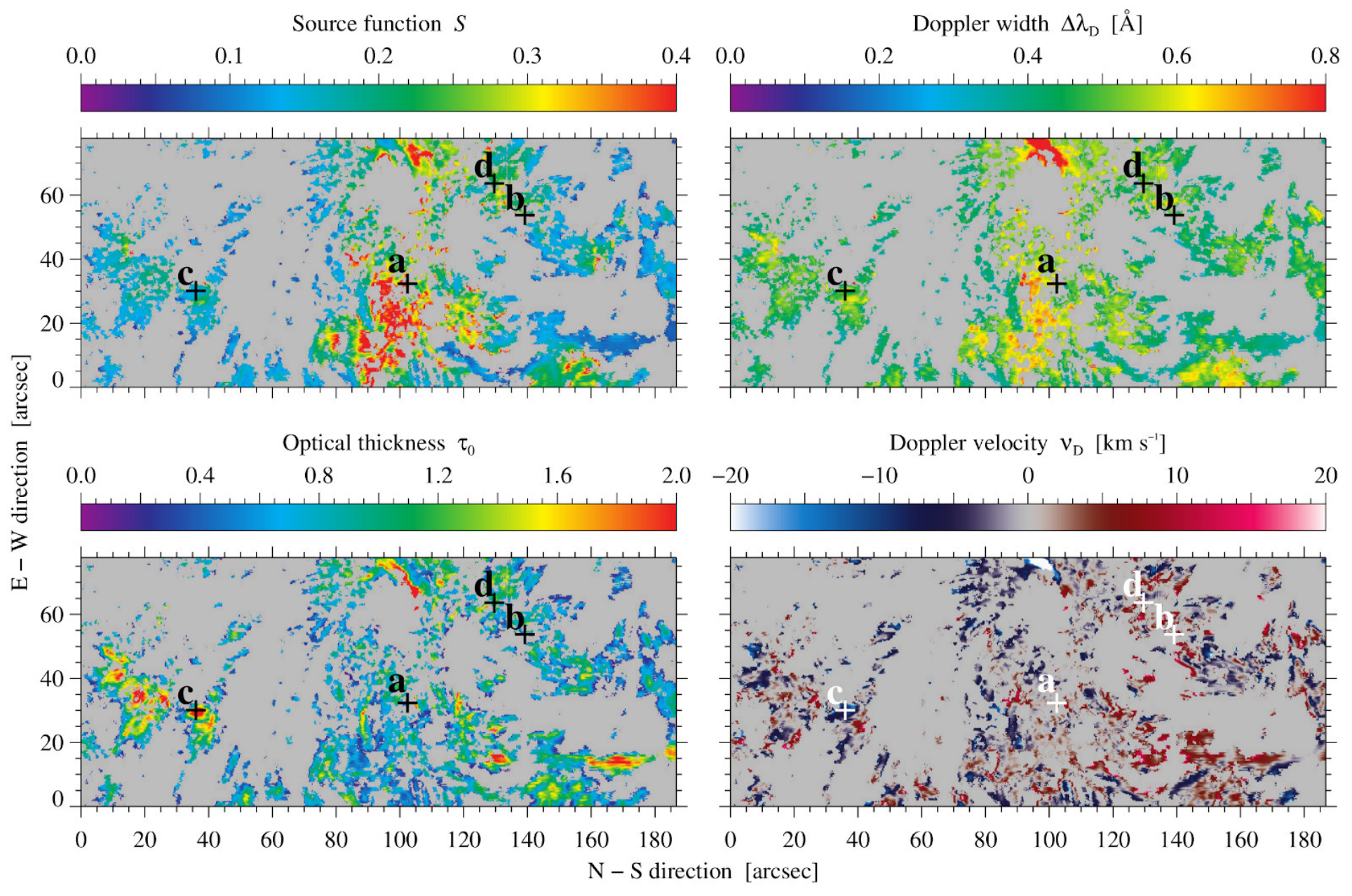}
\caption{Two-dimensional maps of the inversion results for the four CM 
    parameters based on the noise-stripped contrast profiles: source 
    function $S$, Doppler width $\Delta\lambda_\mathrm{D}$ of the absorption 
    profile, optical thickness $\tau_0$, and Doppler velocity of the cloud 
    material $v_\mathrm{D}$ (\textit{from top-left to bottom-right}). The FOV 
    is the same as in Fig.~\ref{FIG01}\ns. Note that the velocity of the cloud 
    material $v_\mathrm{D}$ differs from those depicted in Fig.~\ref{FIG08}\ns.
    Regions, where the CM is unsuitable and inversions fail, are reproduced 
    in gray. The crosses indicate the locations of the profiles in
    Fig.~\ref{FIG07}\ns.}
\label{FIG09}
\end{figure*}

Four sample intensity and contrast profiles are displayed in the left and right panels of Fig.~\ref{FIG07}\ns, respectively. The observed H$\alpha$ intensity profiles still contain blends by telluric and photospheric spectral lines but their signatures are significantly reduced in the contrast profiles. Noise-stripping was based on the ten eigenvectors derived from the CM database and with the appropriate quiet-Sun background interpolated profile for $\mu = 0.81$ according to \citet{David1961}. The noise-free H$\alpha$ contrast profiles were the input of the CM inversions. The selection includes red- and blue-shifted as well as broad and narrow profiles. Direct comparison of the left and right panels of Fig.~\ref{FIG07}{\ns} clearly demonstrates why CM inversions based on contrast profiles are the better choice than intensity profiles. The reason is that the physics of the spectral line formation is encapsulated in the variation around the ``mean'', i.e., the quiet-Sun background profile. Whereas the intensity profiles differ only minutely, i.e., differences in line width and shift are barely recognizable, the contrast profiles exhibit significant variations. After some training, even plain visual inspection of the contrast profiles reveals some of the underlying physics. In general, the fitted profiles are very well represented by the eigenvectors and simply restored using SVD. This is even partly true for contrast profiles, which do not adhere to the CM, because the linear combination of the eigenvectors covers a larger range of contrast profiles compared to those contained in the CM database.

Just with spectral line fitting much information can be extracted from the noise-free line profiles. For example, fitting the line core with a $2^\mathrm{nd}$- or $4^\mathrm{th}$-order polynomials yields the Doppler velocity reflecting plasma motions in the highest chromospheric layers. The strong telluric line contamination close to the H$\alpha$ would have made this task challenging. Bisector analyses delivers information about lower layers. Counterposing line-core and bisector velocity fields in  Fig.~\ref{FIG08}{\ns} exposes significant differences in the flow speeds and the fine structures in both atmospheric layers. Note that because of the complex formation of the H$\alpha$ line, assigning specific heights in the atmosphere to the Doppler maps may be ill-advised. However, already a general trend of the height-dependent flow fields provides insight into the nature of chromospheric absorption features.

The results of the CM inversions are summarized in Fig.~\ref{FIG09}\ns, which depicts two-dimensional maps of all four CM parameters based on the noise-stripped contrast profiles. High values of the source function $S$ are encountered in the vicinity of the small pore in the middle of the bottom part of the FOV. This region also shows some of the strongest line-core Doppler velocities (see bottom panel of Fig.~\ref{FIG08}\ns). In addition, the line profiles are broadened at this location as is evident in the map for the Doppler width $\Delta\lambda_\mathrm{D}$. The strongest absorption features occur on the left side of the FOV, at the extended tips of superpenumbral filaments. Other strong absorption features are associated with the active-region filament in the bottom-right quadrant of the FOV. In general, the cloud velocities $v_\mathrm{D}$ have the same morphology as the two velocity measurements in Fig.~\ref{FIG08}\ns. However, the range covered by the cloud velocities is more than two times larger.

The underlying assumption of the CM is that cool absorbing plasma is suspended by the magnetic field higher up in the solar atmosphere, i.e., regions with enhanced line-core and line-wing emission cannot be expressed by CM inversions. If H$\alpha$ intensity profiles are close to the quiet-Sun profile of the background radiation, CM inversions are also not suitable. The regions are shown in light gray in Fig.~\ref{FIG09}{\ns} and cover about 71.4\% of the observed FOV for the CM inversions based on the noise-stripped contrast profiles. In comparison, this fraction increases to 77.3\% for CM inversions based on the observed contrast profiles. These fractions will, however, change depending on the observed scene on the Sun, i.e., with respect to the coverage of cloud-like features that contain cool, absorbing plasma.

\citet{Verma2012} followed the evolution of decaying active region NOAA~11126 
over the course of five days, and only the data from the first day is discussed 
in the present work. The earlier study focused only on LOS velocity and 
line-core intensity observed in the chromospheric H$\alpha$ line. In the 
H$\alpha$ line-core map, a superpenumbra-like structure was noticed in one of 
the decaying spots, which is also clearly visible in the line-core map in bottom 
the panel of Fig.~\ref{FIG01}\ns. Note that the FOV in Fig.~\ref{FIG01}{\ns} 
covers a larger area compared to \citet{Verma2012} and has a different 
orientation. The larger FOV makes it possible to examine the surroundings of the 
decaying spot. Several interesting features, which were not covered in the 
previous study, are now included, e.g., the long filament in bottom-right quadrant of 
Fig.~\ref{FIG01}{\ns} and the dark surge-like feature associated with the larger 
sunspot. The noise-stripped spectra facilitated computing LOS velocity maps at 
various positions (bisectors) in the H$\alpha$ line profile as shown in 
Fig.~\ref{FIG08}\ns, which was not possible in the previous study and now gives 
access to the height dependence of chromospheric Doppler velocities. Inspecting 
the first map in the middle panel of Fig.~8 in \citet{Verma2012} and 
Fig.~\ref{FIG08}{\ns} of the present work reveals that the noise-stripped 
spectra show more and finer details across a wide range of plasma LOS 
velocities. Note the different velocity ranges of $\pm$5~km~s$^{-1}$ in Fig.~8 
of \citet{Verma2012} and $\pm$8~km~s$^{-1}$ in Fig.~\ref{FIG08}{\ns} of the 
present work. Furthermore, now all four CM parameters (Fig.~\ref{FIG09}\ns) are 
at hand, which provide further means to explore plasma properties, which were 
not computed and discussed in \citet{Verma2012}.

\begin{figure*}[t]
\includegraphics[width=\textwidth]{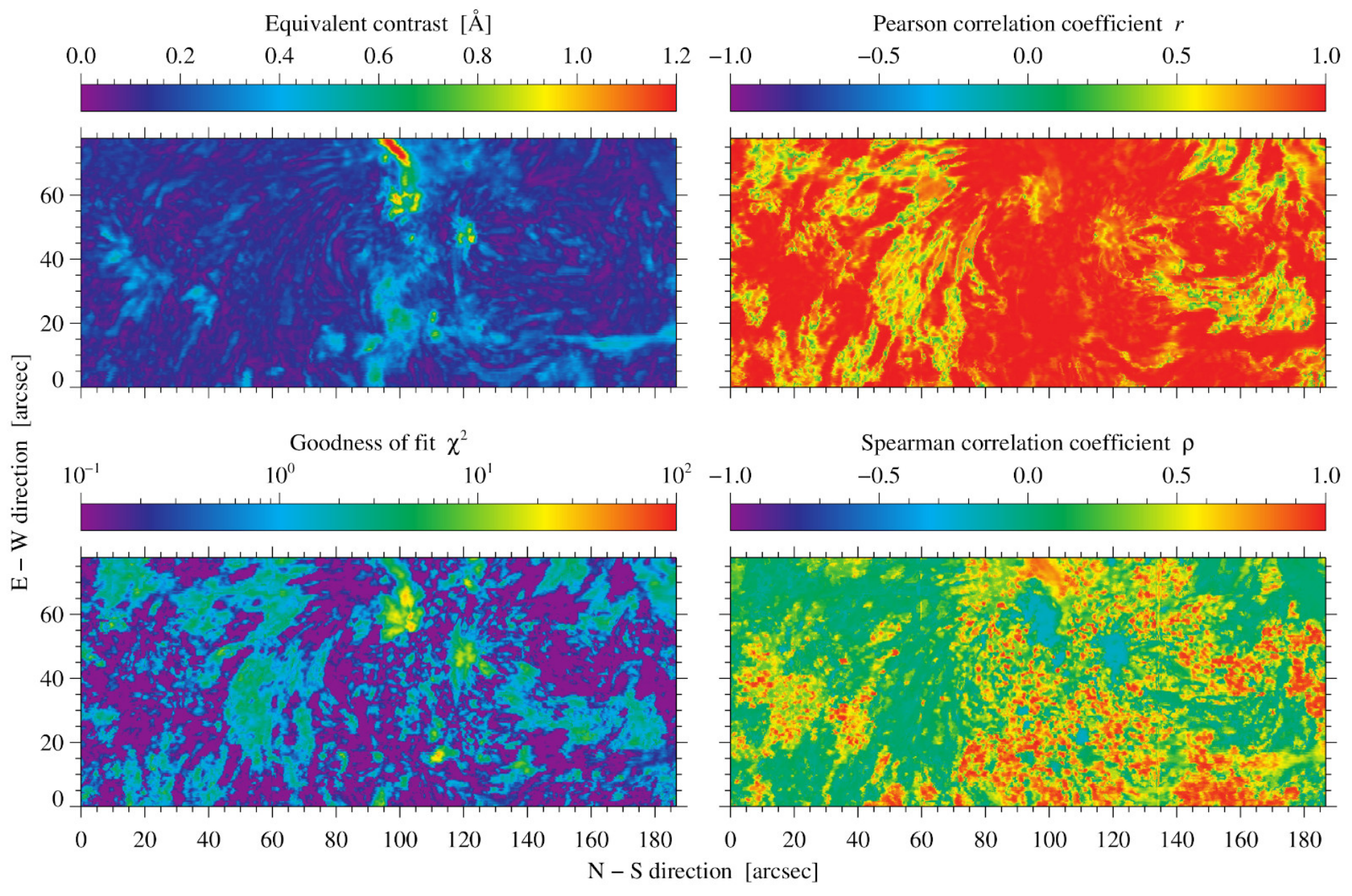}
\caption{Two-dimensional maps illustrating the performance of the CM 
    inversions based on the noise-stripped contrast profiles: 
    equivalent contrast, Pearson's linear correlation coefficient 
    $r$, goodness of fit parameter $\chi^2$, and Spearman's rank-order 
    correlation coefficient $\rho$ (\textit{from top-left to bottom-right}). 
    Note that the goodness of fit parameter $\chi^2$ is depicted on a
    logarithmic scale.}
\label{FIG10}
\end{figure*}

The limitations of CM inversions are addressed in Fig.~\ref{FIG10}\ns. Some of the strongest contrasts are encountered in regions with H$\alpha$ line-core brigthenings (see upper-left panel of Fig.~\ref{FIG10}\ns) or even in emission features, which are both not accessible by the CM inversions. Already the map of the linear correlation coefficient $r$ indicates that the CM is not appropriate for quiet-Sun regions. This finding is seconded by the rank-order correlation coefficient $\rho$, which shows the same morphology but at much lower correlation values. Visual inspection of these two correlation maps led to the above definition of good fits between observed contrast profiles after noise-stripping and those derived from CM inversions. The goodness of the fits is given on a logarithmic scale as SSQ, i.e., the $\chi^2$ statistics, which is over much of the FOV anti-correlated with the linear and rank-order correlation maps. In summary, judging the quality of the fits has always a subjective component, which however can be quantified facilitating the comparison of diverse datasets from different telescopes, instruments, and detectors.

Potential error sources of CM inversions include the determination of the quiet-Sun background profile \citep{Bostanci2010}, photon and detector noise, and numerical precision of the fitting algorithm. The last error source can be neglected if the algorithm does not get trapped in the wrong local minimum, and the second error source is small compared to systematic errors. Beyond the choice of the background profile, the entire data calibration and processing pipeline plays an important role. In addition, mediocre seeing may scramble the spectral information, and spatial resolution dictates how much of the chromospheric fine structure is resolved.

The number of successful CM inversions differs significantly for the observed and noise-stripped contrast profiles (see above). Thus, the comparison was carried out only for the intersection of contrast profiles, i.e., 19.5\% of the contrast profiles. Pearson's linear correlation coefficient is very high, i.e., $\rho = 0.917$ for the optical thickness $\tau_0$ and $\rho \approx 0.99$ for all other CM parameters. Spearman's rank-order correlation coefficient of $r_s = 0.926$ is slightly higher for the optical thickness $\tau_0$ but remains essentially the same for all other CM parameters. This indicates that the two methods capture broadly the physical parameters of the decaying active region NOAA~11126. The mean absolute difference is a quantitative indicator of the systematic error between two methods. It amounts to 0.005 for the source function $S$, to 0.024~\AA\ for the Doppler width $\Delta\lambda_\mathrm{D}$, to 0.103 for the optical thickness $\tau_0$, and to 0.25~km~s$^{-1}$ for the Doppler velocity $v_\mathrm{D}$. The absolute mean differences are comparable to the increments that were used to create the CM database in Sect.~\ref{SEC41}. The mean relative absolute differences are 4.5\%, 4.7\%, 11.9\%, and 18.0\% for the source function $S$, Doppler width $\Delta\lambda_\mathrm{D}$, optical thickness $\tau_0$, and Doppler velocity $v_\mathrm{D}$, respectively. The mean was taken after computing the relative absolute differences for individual data points, avoiding the division by the near-zero arithmetic mean of the Doppler velocity $\bar{v}_\mathrm{D}$. In any case, relative differences are always biased when dividing by small values. In addition, mean value and standard deviation are given for the distributions of the four CM parameters, which serves as an orientation for interpreting errors, i.e., 0.194$\pm$0.088, 0.50$\pm$0.12~\AA, 0.82$\pm$0.41, $-$0.64$\pm$5.52~km~s$^{-1}$, respectively, whereby the standard deviation reflects the variation of the parameters and does not refer to any type of error estimate.


\section{Summary and conclusions}\label{SEC6}

The application of PCA and CM inversions for analyzing the strong chromospheric 
H$\alpha$ line, as presented here, is not the first of its kind. A number of 
studies already demonstrated the robustness of these methods individually 
\citep[e.g.,][]{Rees2000, Tziotziou2007}. However, the intention of the the 
present work was a comprehensive and careful description of the preparation of 
data obtained with the VTT echelle spectrograph: starting with basic 
calibrations, over preprocessing of spatio-spectral data cubes and application 
of PCA for noise-stripping and CM inversions, to finally science-ready data. One
design principle of the data processing pipeline was minimal user interaction 
and robust operations covering the full variety of absorption features in the
solar chromosphere. Admittedly, the CM inversions are computationally intense.
However, already noise-stripping based on PCA and simple line fitting yield 
maps of physical parameters, which provide a quick-look overview of the
dynamic solar atmosphere. The migration of data observed in the past and more
standardized and streamlined data in the future to data archives motivated 
this work. Indeed, the data processing pipeline was successfully tested on 
several VTT datasets. \citet{Verma2019a} presented high-resolution 
H$\alpha$ spectroscopy of active region NOAA~12722, investigating the temporal
evolution of a pore, whereas \citet{Verma2019b} use machine learning and
statistical techniques to classify H$\alpha$ spectra according to results of 
CM inversions. The collaborative research environment and GREGOR 
archive\footnote{\href{https:\\gegor.aip.de}{gregor.aip.de}} at AIP provides the 
solar physics community already with access to GFPI and HiFI data. The goal is 
to integrate AIP's VTT echelle data to the same framework.    

The thermal and dynamic 
properties of various solar features are reflected differently in the H$\alpha$ 
line. For example, \citet{Verma2012} and \citet{Kuckein2016} demonstrate that 
H$\alpha$ spectral scans are still a very useful tool to study the diversity of 
solar features from the evolution of sunspots to the stability of large-scale 
quiescent filaments. In an effort to utilize VTT spectra efficiently, the data 
processing pipeline was critically reviewed, which led to a redesign of the 
data processing and analysis software. Taking advantage of PCA's ability to 
reduce the dimensionality of the CM database and of its capacity for 
noise-stripping, a CM inversion scheme was developed that utilizes only ten 
eigenvectors to reliably reproduce observed lines and robustly delivers the four 
CM parameters. In addition, various steps to calibrate spectral data, which 
are usually only mentioned in passing when presenting the results, are 
discussed in detail because their impact on the inversion result is 
significant. The impact of data (pre)processing on inversion results holds 
certainly true also for other spectral line inversion schemes. Furthermore, 
embedding the CLV of quiet-Sun H$\alpha$ background profiles \citep{David1961} 
facilitated the construction of an extensive database for modeled H$\alpha$ 
intensity and contrast profiles. As a result, not only noise was efficiently 
removed but also the blends of telluric and solar spectral lines in the wings 
and core of the H$\alpha$ line. Thus, computing bisectors becomes 
straightforward providing insight in the height dependence of chromospheric 
velocity fields. 

Comprehensive databases and archives of the H$\alpha$ spectral observations 
do not exist. Despite the importance of H$\alpha$ spectroscopy, no space mission 
carried an H$\alpha$ spectrograph so far but the Narrowband Filter Imager 
(NFI) on board the Japanese \textit{Hinode} mission \citep{Tsuneta2008} had some
Doppler capabilities. However, the Solar H$\alpha$ Imaging 
Spectrometer (SHIS) is considered for an upcoming Chinese space mission. Thus, 
making existing VTT H$\alpha$ data more accessible fills a niche until new 
missions and instruments become available. The streamlined data pipeline for 
bulk processing and analysis of the H$\alpha$ spectra is a first step. Over
the years, a large volume of spatio-spectral data cubes was collected at the
VTT. More recently, an improved observing scheme to capture spectra with a 
faster cadence was implemented at VTT \citep{Denker2019}. 
In the short period of one month more than a billion ($10^9$) individual 
spectral H$\alpha$ profiles were recorded. These data were accompanied by 
chromospheric H$\beta$ spectra and spectra of a photospheric Cr\,\textsc{i} 
line, which has a large Land\'e factor so that magnetic field information can be 
glimpsed from intensity spectra. The data processing pipeline will be adapted to 
these two lines as well. In the next step, other chromospheric lines (e.g., 
Ca\,\textsc{ii}\,H\,\&\,K, Na\,\textsc{i}\,D$_1$\,\&\,D$_2$, and near-infrared 
Ca\,\textsc{ii}), which were also observed with the VTT in the past, will be 
made accessible, too.


\section{Acknowledgments}

The  Vacuum  Tower  Telescope  at  the  Spanish  Observatorio  del  Teide  of  
the Instituto  de Astrof\'{\i}sica  de Canarias is operated by the German  
consortium  of the Leibniz-Institut f\"ur Sonnenphysik in Freiburg, the 
Leibniz-Institut f\"ur Astrophysik Potsdam, and the Max-Planck-Institut f\"ur 
Sonnensystemforschung G\"ottingen. This study was supported by grant DE~787/5-1 
of the Deutsche Forschungsgemeinschaft (DFG) and by the European Commission's 
Horizon 2020 Program under grant agreements 824064 (ESCAPE -- European 
Science Cluster of Astronomy \& Particle physics ESFRI research infrastructures) 
and 824135 (SOLARNET -- Integrating High Resolution Solar Physics). ED is 
grateful for the generous financial support from German Academic Exchange 
Service (DAAD) in form of a doctoral scholarship. SJGM and PS acknowledge the 
support of the project VEGA 2/0004/16. We would like to thank the referee
who provided helpful comments and guidance, improving structure and contents
of the manuscript.
\clearpage


\end{document}